\documentclass[twocolumn,showpacs,preprintnumbers,amsmath,amssymb]{revtex4}
\usepackage{epsfig}
\usepackage{dcolumn}
\usepackage{bm}

\newcommand{\remove}[1]{}

\def\be{\begin{equation}}
\def\ee{\end{equation}}

\newcommand{\beq}{\begin{equation}}
\newcommand{\eeq}{\end{equation}}
\newcommand{\beqa}{\begin{eqnarray}}
\newcommand{\eeqa}{\end{eqnarray}}

\newcommand{\ii}{{\rm i}}

\newcommand{\vv}{{\bf v}}

\newcommand{\vx}{{\bf x}}
\newcommand{\vk}{{\bf k}}
\newcommand{\vp}{{\bf p}}
\newcommand{\vq}{{\bf q}}

\newcommand{\cD}{{\cal D}}

\newcommand{\cP}{{\cal P}}

\newcommand{\bea}{\begin{array}}
\newcommand{\ea}{\end{array}}

\begin{document}

\title{Consistency relations for large-scale structures with primordial non-Gaussianities}

\author{Patrick Valageas$^{1,2}$, Atsushi Taruya$^{3,4}$, Takahiro Nishimichi$^{4,5}$}

\affiliation{
$^1$Institut de Physique Th\'eorique,
CEA, IPhT, F-91191 Gif-sur-Yvette, C\'edex, France\\
$^2$CNRS, URA 2306, F-91191 Gif-sur-Yvette, C\'edex, France\\
$^3$Center for Gravitational Physics, Yukawa Institute for Theoretical Physics, Kyoto University, Kyoto 606-8502, Japan\\
$^4$Kavli Institute for the Physics and Mathematics of the Universe, Todai Institutes for Advanced Study, the University of Tokyo, Kashiwa, Chiba 277-8583, Japan (Kavli IPMU, WPI)\\
$^5$CREST, JST, 4-1-8 Honcho, Kawaguchi, Saitama, 332-0012, Japan}

\date{\today}
\vspace{.2 cm}

\begin{abstract}

We investigate how the consistency relations of large-scale structures are modified
when the initial density field is not Gaussian.
We consider both scenarios where the primordial density field can be written as 
a nonlinear functional of a Gaussian field and more general scenarios where 
the probability distribution of the primordial density field can be expanded around the
Gaussian distribution, up to all orders over $\delta_{L0}$.
Working at linear order over the non-Gaussianity parameters $f_{\rm NL}^{(n)}$ or
$S_n$, we find that the consistency relations for the matter density fields are modified as they include additional
contributions that involve all-order mixed linear-nonlinear correlations 
$\langle \prod \delta_L \prod \delta \rangle$.
We derive the conditions needed to recover the simple Gaussian form of the consistency
relations. This corresponds to scenarios that become Gaussian in the squeezed limit.
Our results also apply to biased tracers, and velocity or momentum cross-correlations.

\end{abstract}

 \vskip2pc

\maketitle

\section{Introduction}
\label{introduction}

Large-scale matter inhomogeneities in the Universe, as partly probed by luminous distributions such as 
galaxies and clusters, are thought to have evolved from tiny density fluctuations under the influence of 
both gravity and cosmic expansion. Since the observations of large-scale structures are basically made with 
measurements of both redshift and angular positions, the three-dimensional matter distribution contains valuable 
cosmological information, complementary to the cosmic microwave background observed in a two-dimensional sky. 
With the precision measurements provided by the ongoing and future surveys, the large-scale structure 
observations may also give us hints and clues beyond the standard cosmological model. 

While the evolution of matter inhomogeneities until the time of decoupling is fully described by the linear 
theory, based on the cosmological Boltzmann equations coupled with Einstein's theory of general relativity 
(e.g., \cite{Lewis2000}), the development of gravitational clustering becomes significant at late time, 
and it eventually enters the nonlinear stage, where the applicability of analytic predictions is severely restricted 
and only partly accessible through perturbation theory \cite{Bernardeau2002}. 
In general, numerical simulations are required for quantitative cosmological studies. 

However, it has been recently recognized that there exist nonperturbative statistical relationships that hold 
even in the nonlinear regime. These so-called {\it consistency relations} represent a nontrivial coupling between 
large- and small-scale modes, and lead to exact relationships between higher- and lower-order statistics 
of density fluctuations (e.g., \cite{Kehagias2013,Peloso2013,Peloso2014, Creminelli2013,Kehagias2014c,Creminelli2014a,Valageas2014a,Horn2014,Horn2015}). 
They generically hold not only for the density fluctuations of the matter distribution but also 
for biased tracers. 
The original consistency relations are kinematic and vanish at equal times at leading order.
One can also take into account next-to-leading-order contributions to obtain relations that remain
nonzero at equal times \cite{Valageas2014b,Kehagias2014,Nishimichi2014}, but this requires some
additional approximations so that these relations are no longer exact.
On the other hand, by considering cross-correlations between density and velocity, or momentum, fields,
one can again derive exact relations, analogs of the original density relations, that remain nonzero at equal times
\cite{Rizzo2016a} and are potentially important as they may be directly measurable via observations. 
Since these exact consistency relations only rely on the weak equivalence principle and the Gaussianity of 
the primordial fluctuations, their measurement could be a powerful test of the fundamental hypothesis 
of the standard cosmological model. 

A few works have considered how these relations are modified when one of the underlying assumptions 
is not fulfilled, namely if we have violations of the weak equivalence principle in modified-gravity scenarios
\cite{Kehagias2014c,Peloso2014,Creminelli2014}, or primordial non-Gaussianities
\cite{Peloso2013}, by studying a few examples.
This aspect is crucial if we wish to test the standard cosmological model through the 
measurement of these consistency relations. 
In this paper, we investigate in more details how the consistency relations are modified in the presence 
of primordial non-Gaussianity, considering both very general non-Gaussian models and all-order consistency
relations.

The primordial non-Gaussianity is now severely constrained through the measurement 
of cosmic microwave background anisotropies \cite{2013arXiv1303.5084P,Ade:2015ava}, 
but the outcome of these tight constraints relies on several model-dependent assumptions. 
This is one of the reasons why the large-scale structure observations still attract attention 
as an independent and complementary probe of primordial non-Gaussianity. A particularly remarkable feature 
that has been recently recognized is a strong enhancement of the halo and galaxy clustering bias on large scales 
in the presence of the so-called local-type non-Gaussianity (e.g., \cite{Dalal2008,Afshordi2008,Slosar:2008hx}). 
This enhancement indeed arises from a tight coupling between the large- and small-scale modes through the 
squeezed limit of higher-order matter correlations (e.g., \cite{2012PhRvD..86f3518M}), which is exactly the case 
we are looking at in the consistency relations. It is thus interesting to see how the structure of the correlation 
hierarchy is generically modified when the initial fluctuations are not Gaussian. 

Hereafter, we consider non-Gaussian primordial matter fluctuations described by their Taylor expansion 
over a Gaussian field, or by a probability distribution that can be expanded around the Gaussian.
We derive rather generic expressions for the higher-order correlations of matter fluctuations that remain valid 
in the nonlinear regime. We also consider biased tracers.
Our results show that in the non-Gaussian case the consistency relations are largely 
modified by additional contributions involving all-order mixed linear-nonlinear correlations. 
Based on this, we discuss the necessary conditions to recover the usual consistency relations that hold 
for Gaussian initial conditions, and see how these conditions are satisfied or violated in specific 
non-Gaussian models. 

This paper is organized as follows.
In Sec.~\ref{sec:model-NG}, we begin by describing non-Gaussian primordial density fields as Taylor expansions 
of an auxiliary Gaussian field. Next, we consider the more general case where the primordial density field is
merely defined by its non-Gaussian probability distribution, which we assume can be expanded around
the Gaussian. We also recall several popular non-Gaussian models that provide useful examples.
Sec.~\ref{sec:correlation_response} considers the basis to derive the consistency relations \cite{Valageas2014a}, 
and derives the relation between the response functions of cosmic density fields, with respect to the linear density 
field, and higher-order correlation functions. 
Sec.~\ref{sec:density_contrast} then presents our main results, which describe how the consistency relations 
of density fields are generically modified in the presence of primordial non-Gaussianity. 
Specific results for several non-Gaussian models are also given. 
Further, Sec.~\ref{sec:consistency_rel-v-p} discusses the consistency relations for velocity and momentum fields. 
Finally, Sec.~\ref{sec:Conclusions} is devoted to the conclusion and summary of the results.

\section{Models of primordial non-Gaussianities}
\label{sec:model-NG}

In this section, as a starting point to derive the correlation hierarchy in the presence of primordial non-Gaussianity, 
we give a general framework to deal with non-Gaussian matter fluctuations at linear order. 
In Sec.~\ref{sec:NLmapping}, we present a description of non-Gaussian primordial density fields through 
their Taylor expansion over auxiliary Gaussian fields. 
In Sec.~\ref{sec:NLproba}, this is generalized to the probability distribution functional for non-Gaussian
primordial fluctuations, which is later used to derive the consistency relations for matter fluctuations. 
As a simple and illustrative example, in Sec.~\ref{sec:quadratic}, we consider the non-Gaussian model 
in which the Taylor expansion is truncated at second order. Specific models to realize such a non-Gaussianity 
are described in Sec.~\ref{sec:explicit}, and we briefly discuss their distinct features in the squeezed limit.

\subsection{Primordial density field as a nonlinear functional of a Gaussian field}
\label{sec:NLmapping}

Simple models of primordial non-Gaussianities can be built where the primordial (i.e., linear) 
density contrast $\delta_L(\vx,\tau)$ can be written as a nonlinear functional of a Gaussian 
field $\chi(\vx,\tau)$.
Linearizing over the non-Gaussianity parameters $f_{\rm NL}^{(n)}$, as we do throughout
this study, we write in Fourier space
\beqa
\delta_{L0}(\vk) & = & \chi_0(\vk) + \sum_{n=2}^\infty \int \prod_{i=1}^ n d\vk_i \;
\delta_D \! \biggl( \vk - \sum_{i=1}^n \vk_i \biggl) \nonumber \\
&& \times f_{\rm NL 0}^{(n)}(\vk_1,...,\vk_n) \prod_{i=1}^n \chi_0(\vk_i)  ,
\label{deltaL0-chi0}
\eeqa
where the subscript ``0'' denotes that we normalize the fields today at $z=0$
and we can take the kernels $f_{\rm NL}^{(n)}$ to be symmetric.
Throughout this paper we assume statistical homogeneity, hence the Dirac factors
in Eq.(\ref{deltaL0-chi0}), and isotropy, which yields the constraint
\beq
f_{\rm NL}^{(n)}(\vk_1,...,\vk_n) = f_{\rm NL}^{(n)}(-\vk_1,...,-\vk_n) .
\label{fNL-isotropy}
\eeq
For the sake of generality we keep track of all orders $n \geq 2$ in the nonlinear
functional $\delta_L[\chi]$, but in practice one often only includes the quadratic or cubic
terms, which are then denoted as $f_{\rm N}=f_{\rm NL}^{(2)}$ and
$g_{\rm NL}=f_{\rm NL}^{(3)}$.
The fields $\delta_L$, $\chi$, and the kernels $f_{\rm NL}^{(n)}$ evolve with redshift as
\beq
\delta_L = D_+ \delta_{L0} , \;\;\; \chi = D_+ \chi_0 , \;\;\; 
f_{\rm NL}^{(n)} = D_+^{1-n} f_{\rm NL0}^{(n)} ,
\label{D+}
\eeq
where $D_+(\tau)$ is the linear growing mode. The kernels $f_{\rm NL}^{(n)}$ 
must satisfy the constraint
\beq
n \geq 2: \;\;\; f_{\rm NL}^{(n)}(\vk_1,...,\vk_n) = 0 \;\;\; \mbox{for} \;\;\; \vk_1+...+\vk_n = 0 ,
\label{fNL=0}
\eeq
so that $\delta_{L}(0)=0$. At linear order over $f_{\rm NL}^{(n)}$, this yields the primordial 
power spectrum
\beqa
&& \hspace{-1cm} \langle \delta_{L0}(\vk) \delta_{L0}(-\vk) \rangle' \equiv P_{L0}(k) \nonumber \\
&& = P_{\chi_0}(k) + 2 P_{\chi_0}(k) \sum_{n=1}^{\infty} (2n+1)!! 
\int \prod_{i=1}^n d\vk_i'  \nonumber \\
&& \times f^{(2n+1)}_{\rm NL0}(\vk,\vk_1',-\vk_1',...,\vk_n',-\vk_n')  
\prod_{i=1}^n P_{\chi_0}(k_i') 
\label{Pk}
\eeqa
and the primordial bispectrum
\beqa
&& \hspace{-0.5cm} \langle \delta_{L0}(\vk_1) \delta_{L0}(\vk_2) \delta_{L0}(\vk_3) \rangle' 
= P_{\chi_0}(k_2) P_{\chi_0}(k_3) \sum_{n=1}^{\infty} 2n \nonumber \\
&& \times (2n-1)!! \int \prod_{i=1}^{n-1} d\vk_i'  \; \prod_{i=1}^{n-1} P_{\chi_0}(k_i') 
\nonumber \\
&& \times f^{(2n)}_{\rm NL0}(\vk_2,\vk_3,\vk_1',-\vk_1',...,\vk_{n-1}',-\vk_{n-1}') 
+ 2 \; \rm{cyc.} \;\;\;
\label{Bk}
\eeqa
where the prime in $\langle\dots\rangle'$ denotes that we removed the Dirac factors
$\delta_D(\sum\vk_i)$.

Because the relevant field for the formation of large-scale structures is the linear density
field $\delta_{L0}$ rather than the auxiliary Gaussian field $\chi_0$, it is convenient
to eliminate $\chi_0$ in favor of $\delta_{L0}$.  
This is possible because, at linear order over $f_{\rm NL}^{(n)}$, we can invert 
Eq.(\ref{deltaL0-chi0}) as
\beqa
\chi_0(\vk) & = & \delta_{L0}(\vk) - \sum_{n=2}^\infty \int \prod_{i=1}^ n d\vk_i \;
\delta_D \! \biggl( \vk - \sum_{i=1}^n \vk_i \biggl) \nonumber \\
&& \times f_{\rm NL 0}^{(n)}(\vk_1,...,\vk_n) \prod_{i=1}^n \delta_{L0}(\vk_i) 
+ {\cal O}(f_{\rm NL}^2) . \hspace{1cm}
\label{chi0-deltaL0}
\eeqa
The generating functional $\langle e^{j\cdot\delta_{L0}} \rangle$ reads as
\beq
\langle e^{j\cdot\delta_{L0}} \rangle = \int \cD\chi_0 \; e^{j\cdot\delta_{L0}[\chi_0]} \; 
e^{-\chi_0 \cdot C_{\chi_0}^{-1} \cdot \chi_0 /2} ,
\eeq
as $\chi_0$ is Gaussian and we introduced the inverse matrix $C_{\chi_0}^{-1}$ of
the two-point correlation 
$C_{\chi_0}(\vk_1,\vk_2) \equiv \delta_D(\vk_1+\vk_2) P_{\chi_0}(k_1)$.
Using Eq.(\ref{chi0-deltaL0}), we can change the variable to $\delta_{L0}$ to obtain
\beqa
&& \hspace{-0.5cm} \langle e^{j\cdot\delta_{L0}} \rangle = \int \!\! \cD\delta_{L0} \; J \;
\exp \biggl [ j\cdot\delta_{L0}
- ( \delta_{L0} \! - \! \sum_n f_{\rm NL0}^{(n)} \delta_{L0} ... \delta_{L0} ) \;\; \nonumber \\
&& \hspace{1cm} \cdot C_{\chi_0}^{-1} \cdot
( \delta_{L0} - \sum_n f_{\rm NL0}^{(n)} \delta_{L0} ... \delta_{L0} ) /2 \biggl ] ,
\label{partition-1}
\eeqa
where the Jacobian determinant reads at linear order over $f_{\rm NL}^{(n)}$ as
\beqa
J & \equiv & \left| \det\left( \frac{\cD\chi_0}{\cD\delta_{L0}} \right) \right| \nonumber \\
& = & 1 - \sum_{n=3}^{\infty} n \int d\vk \int \prod_{i=1}^{n-1} d\vk_i' \; 
\delta_D \biggl( \sum_{i=1}^{n-1} \vk_i' \biggl) \nonumber \\
&& \times \; f_{\rm NL0}^{(n)} (\vk,\vk_1',...,\vk_{n-1}')
\prod_{i=1}^{n-1} \delta_{L0}(\vk_i')  .
\label{Jacobian}
\eeqa
Here we used the fact that the first term $n=2$ in Eq.(\ref{Jacobian}) vanishes
because $\delta_{L0}(\vk_1'=0)=0$ 
[as enforced by the Gaussian weight with $P_{\chi_0}(0)=0$].
Expanding Eq.(\ref{partition-1}) up to first order over $f_{\rm NL}^{(n)}$ we obtain
\beqa
\langle e^{j\cdot\delta_{L0}} \rangle & = & \int \!\! \cD\delta_{L0} \; 
e^{j\cdot\delta_{L0} - \delta_{L0} \cdot C_{\chi_0}^{-1} \cdot \delta_{L0} /2} \; 
\biggl [ 1 + \sum_{n=2}^{\infty} \int \prod_{i=1}^n d\vk_i \nonumber \\
&& \times \; \delta_D\left( \sum_{i=1}^n \vk_i\right) S_n(\vk_1,...,\vk_n) 
\prod_{i=1}^n \delta_{L0}(\vk_i) \biggl ]
\label{partition-2}
\eeqa
where we introduced the symmetric kernels, for $n \geq 2$,
\beqa
S_n(\vk_1,...,\vk_n) & = & - (n+1) \int d\vk' f_{\rm NL0}^{(n+1)}(\vk',\vk_1,...,\vk_n)
\nonumber \\
&& + \frac{1}{n} \sum_{\rm cyc.} \frac{1}{P_{\chi_0}(k_1)} 
f_{\rm NL0}^{(n-1)}(\vk_2,...,\vk_n) , \hspace{1cm}
\label{Sn-def}
\eeqa
with $f_{\rm NL0}^{(1)}(k) \equiv 0$ and the sum runs over the $n$ cyclic permutations
of $\{\vk_1,...,\vk_n\}$.
This means that the probability distribution functional of the linear density
field $\delta_{L0}$ reads as
\beqa
\cP(\delta_{L0}) & = & e^{- \int \! d\vk \; \delta_{L0}(\vk)\delta_{L0}(-\vk)/2P_{\chi_0}(k)}
\biggl [ 1 + \sum_{n=2}^{\infty} \int \prod_{i=1}^n d\vk_i \nonumber \\
&& \hspace{-1cm} \times \; \delta_D\left( \sum_{i=1}^n \vk_i\right) S_n(\vk_1,...,\vk_n) 
\prod_{i=1}^n \delta_{L0}(\vk_i) \biggl ] .
\label{P-S3}
\eeqa

\subsection{Primordial density field with a non-Gaussian probability distribution functional}
\label{sec:NLproba}

Independently of any nonlinear mapping to an auxiliary Gaussian field $\chi_0$,
as in Eq.(\ref{deltaL0-chi0}), we can define the initial conditions of the cosmological
density field $\delta_{L0}$ by its probability distribution functional $\cP(\delta_{L0})$.
When we go beyond the Gaussian case, we face an infinite number of possibilities;
however, we may consider distributions of the same form as Eq.(\ref{P-S3}).
Here the kernels $S_n$ are no longer given in terms of kernels $f_{\rm NL0}^{(n)}$
as in Eq.(\ref{Sn-def}). They define the probability distribution (\ref{P-S3})
of the initial conditions, through the expansion of its non-Gaussian part over $\delta_{L0}$.
This approach is more general than the explicit models (\ref{deltaL0-chi0}) and 
it also applies to multifield scenarios, where the final density perturbations are produced
by a combination of several primordial fields.

The term $n=1$ in the expansion (\ref{P-S3}) still vanishes because $\delta_{L0}(0)=0$.
The normalization condition $\langle 1 \rangle =1$ implies that the kernels $S_n$
must obey the constraint
\beqa
&& \sum_{n=1}^{\infty} (2n-1)!! \int \prod_{i=1}^n d\vk_i \;
S_{2n}(\vk_1,-\vk_1,...,\vk_n,-\vk_n) \nonumber \\
&& \times \prod_{i=1}^n P_{\chi_0}(k_i)  = 0 ,
\label{Sn-norm-0}
\eeqa
while the condition $\langle \delta_{L0}(\vk) \rangle = 0$ gives the constraint
\beqa
&& \sum_{n=1}^{\infty} (2n+1)!! \int \prod_{i=1}^n d\vk_i \;
S_{2n+1}(0,\vk_1,-\vk_1,...,\vk_n,-\vk_n) \;\;\;  \nonumber \\
&& \times \; P_{\chi_0}(0) \prod_{i=1}^n P_{\chi_0}(k_i)  = 0 .
\label{Sn-norm-1}
\eeqa
We can check that the explicit models (\ref{Sn-def}) satisfy the conditions
(\ref{Sn-norm-0})-(\ref{Sn-norm-1}).
The first condition (\ref{Sn-norm-0}) is satisfied because of the cancellation
in the sum over $n$ between the two terms in Eq.(\ref{Sn-def}).
The second condition (\ref{Sn-norm-1}) is satisfied because $P_{\chi_0}(0)=0$
and $f_{\rm NL0}^{(2n)}(\vk_1,-\vk_1,...,\vk_n,-\vk_n)=0$ from 
Eq.(\ref{fNL=0}). As expected, it involves the constraints on $P_{\chi_0}$
and $f_{\rm NL0}^{(n)}$ found in section~\ref{sec:NLmapping} that are
associated with the condition $\delta_L(0)=0$.
[The property $P_{\chi_0}(0)=0$ is not sufficient to ensure Eq.(\ref{Sn-norm-1})
because of a factor $1/P_{\chi_0}(0)$ that arises from the second term in
Eq.(\ref{Sn-def}).]

Then, the linear density power spectrum reads as
\beqa
&& \hspace{-0.5cm} P_{L0}(k) \equiv \langle \delta_{L0}(\vk) \delta_{L0}(-\vk) \rangle' 
= P_{\chi_0}(k) + P_{\chi_0}(k)^2  \nonumber \\
&& \times \sum_{n=1}^{\infty} 2n \, (2n-1)!!
\int \prod_{i=1}^{n-1} d\vk_i' \; \prod_{i=1}^{n-1} P_{\chi_0}(k_i') \nonumber \\
&& \times \; S_{2n}(\vk,-\vk,\vk_1',-\vk_1',...,\vk_{n-1}',-\vk_{n-1}') ,
\label{Pk-Sn}
\eeqa
while the primordial density bispectrum reads as
\beqa
&& \hspace{-0.5cm} \langle \delta_{L0}(\vk_1) \delta_{L0}(\vk_2) \delta_{L0}(\vk_3) \rangle' 
= P_{\chi_0}(k_1) P_{\chi_0}(k_2) P_{\chi_0}(k_3) \nonumber \\
&& \times \sum_{n=1}^{\infty} 2n \, (2n+1)!!
\int \prod_{i=1}^{n-1} d\vk_i' \; \prod_{i=1}^{n-1} P_{\chi_0}(k_i') \nonumber \\
&& \times \; S_{2n+1}(\vk_1,\vk_2,\vk_3,\vk_1',-\vk_1',...,\vk_{n-1}',-\vk_{n-1}') .
\label{Bk-Sn}
\eeqa
In the case of the generalized local models (\ref{deltaL0-chi0}), we can check from
Eq.(\ref{Sn-def}) that we recover the expressions (\ref{Pk}) and (\ref{Bk}).

\subsection{$S_3$-type primordial non-Gaussianity}
\label{sec:quadratic}

As a simple example of non-Gaussian models, let us consider the case of quadratic models, 
where the expansion (\ref{deltaL0-chi0}) stops at the quadratic term,
\beqa
\delta_{L0}(\vk) & = & \chi_0(\vk) + \int d\vk_1 d\vk_2 \;
\delta_D( \vk - \vk_1-\vk_2) \nonumber \\
&& \times f_{\rm NL 0}(\vk_1,\vk_2) \chi_0(\vk_1) \chi_0(\vk_2)  ,
\label{deltaL0-chi0-quadratic}
\eeqa
that is,
\beq
f_{\rm NL}^{(2)}(\vk_1,\vk_2) = f_{\rm NL}(\vk_1,\vk_2), \;\;\;
f_{\rm NL}^{(n)} = 0 \;\; \mbox{for} \;\; n \geq 3.
\eeq
This includes, for instance, the local model described in 
Eqs.(\ref{local-Phi})-(\ref{fNL-local}) below.
Then, the kernels $S_n$ introduced in Eq.(\ref{Sn-def}) are
\beq
S_3(\vk_1,\vk_2,\vk_3) = \frac{f_{\rm NL0}(\vk_2,\vk_3)}{3P_{\chi_0}(k_1)} 
+ 2 \; {\rm cyc.}
\label{S3-f2-quadratic}
\eeq
and
\beq
S_n = 0 \;\; \mbox{for} \;\; n \neq 3.
\eeq
The normalization constraint (\ref{Sn-norm-0}) is trivially satisfied as it reads $0=0$.
The condition (\ref{Sn-norm-1}) associated with the constraint 
$\langle \delta_{L0}(\vk) \rangle = 0$ reads as
\beq
\int d\vk \; S_3(0,\vk,-\vk) P_{\chi_0}(0) P_{\chi_0}(k) = 0 .
\label{S3-zero-mean}
\eeq
From Eq.(\ref{S3-f2-quadratic}) this reads as
\beq
\int d\vk f_{\rm NL0}(\vk,-\vk) P_{\chi_0}(k) + 2 \int d\vk f_{\rm NL0}(0,\vk) P_{\chi_0}(0) = 0 ,
\label{fNL-zero-mean}
\eeq
which is satisfied as both terms vanish, thanks to Eq.(\ref{fNL=0}) and $P_{\chi_0}(0)=0$.

The power spectrum (\ref{Pk-Sn}) is simply
\beq
P_{L0}(k) = P_{\chi_0}(k) ,
\label{PL0-S3}
\eeq
while the bispectrum (\ref{Bk-Sn}) reads as
\beqa
\langle \delta_{L0}(\vk_1) \delta_{L0}(\vk_2) \delta_{L0}(\vk_3) \rangle' 
& = & P_{\chi_0}(k_1) P_{\chi_0}(k_2) P_{\chi_0}(k_3) \nonumber \\
&& \times 6 \; S_3(\vk_1,\vk_2,\vk_3) .
\label{Bk-S3}
\eeqa
Using Eq.(\ref{S3-f2-quadratic}), we recover the usual result
\beqa
\langle \delta_{L0}(\vk_1) \delta_{L0}(\vk_2) \delta_{L0}(\vk_3) \rangle' 
& = & 2 f_{\rm NL0}(\vk_2,\vk_3) \nonumber \\
&& \hspace{-1cm}  \times P_{\chi_0}(k_2) P_{\chi_0}(k_3) + 2 \; {\rm cyc.} 
\label{bispectrum-deltaL0-fNL}
\hspace{1cm}
\eeqa

More generally, beyond the explicit quadratic models (\ref{deltaL0-chi0-quadratic}),
we can define non-Gaussian models by the kernel $S_3$ itself, without reference
to an auxiliary Gaussian field $\chi_0$.
With $S_n=0$ for $n \neq 3$, this provides the simplest probability distribution
${\cal P}(\delta_{L0})$ of Eq.(\ref{P-S3}), which defines the initial conditions for the
density field, that is fully determined by the power spectrum and the bispectrum.
Therefore, this can be seen as the simplest model (in terms of the distribution
function) of a non-Gaussian primordial density field when we only know its second and 
third-order moments.
Then, the kernel $S_3$ simply needs to satisfy the condition (\ref{S3-zero-mean}) to provide
a physical model.

\subsection{Explicit examples of non-Gaussian models}
\label{sec:explicit}

\subsubsection{Local model}
\label{sec:local}

A specific example of such $S_3$-type primordial non-Gaussianity is the quadratic 
local model, where we write Bardeen's potential $\Phi$ as
\beq
\mbox{local type:} \;\;\; \Phi(\vx) = \phi(\vx) + f_{\rm NL} \left( \phi(\vx)^2 
- \langle \phi^2 \rangle \right) ,
\label{local-Phi}
\eeq
where $\phi$ is a Gaussian field and $f_{\rm NL}$ a parameter.
On subhorizon scales, the Poisson equation gives
\beq
\delta_L(\vk,\tau) = \alpha(k,\tau) \Phi(\vk)  \;\; \mbox{with} \;\;
\alpha(k,\tau) = \frac{2c^2k^2T(k)D_+(\tau)}{3\Omega_{\rm m0} H_0^2} ,
\label{Poisson-transfer}
\eeq
where $T(k)$ is the transfer function.
Then, defining $\chi(\vk,\tau) = \alpha(k,\tau) \phi(\vk)$, we obtain the relation 
(\ref{deltaL0-chi0-quadratic}) with
\beq
\mbox{local type:} \;\;\; f_{\rm NL}(\vk_1,\vk_2;\tau) = f_{\rm NL} 
\frac{\alpha(\vk_1+\vk_2,\tau)}{\alpha(k_1,\tau)\alpha(k_2,\tau)} .
\label{fNL-local}
\eeq
The general form (\ref{deltaL0-chi0}) can describe a more general scale dependence
of the primordial non-Gaussianity than the specific kernel (\ref{fNL-local}) and higher-order
contributions.

\subsubsection{Equilateral and orthogonal models}
\label{sec:equilateral_orthogonal}

Other than the local-type primordial non-Gaussianity, there are models discussed in the literature that 
produce distinctive features in the primordial bispectrum. The so-called equilateral- and orthogonal-type 
non-Gaussianities are known to characterize a possible deviation from the simplest single-field inflation scenario 
(e.g., \cite{Chen:2006nt,Senatore:2009gt}). 
While the primordial bispectrum of the former type has peaks at equilateral configurations,  the latter type gives 
a signal peaked both on equilateral and flat-triangle configurations but with opposite sign. 
The bispectrum of these models can be expressed by the template \cite{Scoccimarro:2011pz}
\beqa
&& \langle \Phi(\vk_1) \Phi(\vk_2) \Phi(\vk_3) \rangle' = 6 \, f_{\rm NL}
\nonumber\\
&& \times \Bigl \{ c_1 \left[ P_\Phi(k_1) P_\Phi(k_2) + 2 \; \rm{cyc.} \right]
\nonumber\\
&& \;\;\; + c_2 \left[ P_\Phi(k_1)^{1/3} P_\Phi(k_2)^{2/3} P_\Phi(k_3) + 5 \; \rm{cyc.} \right]
\nonumber\\
&& \;\;\; + c_3 \left[ P_\Phi(k_1) P_\Phi(k_2) P_\Phi(k_3) \right]^{2/3} \Bigr \}
\label{Bispectrum-Phi-c}
\eeqa
with the coefficients being $(c_1,c_2,c_3)=(-1,1,-2)$ for equilateral type and $(-3,3,-8)$ for orthogonal type. 
Note that with the expression given above, the local model (\ref{local-Phi}) corresponds to 
$(c_1,c_2,c_3)=(1/3,0,0)$. 
Using the relation (\ref{Poisson-transfer}) between the primordial gravitational potential $\Phi$ and the
matter-era linear density contrast $\delta_L$, Eq.(\ref{Bispectrum-Phi-c}) gives the linear matter density bispectrum
\beqa
&& \langle \delta_L(\vk_1) \delta_L(\vk_2) \delta_L(\vk_3) \rangle' = 6 \, f_{\rm NL} 
\nonumber\\
&& \times \Biggl \{ c_1 \left[ \frac{\alpha(k_3) P_L(k_1) P_L(k_2)}{\alpha(k_1) \alpha(k_2)} 
+ 2 \; \rm{cyc.} \right] \nonumber\\
&& \;\;\; + c_2 \left[ \frac{\alpha(k_1)^{1/3} P_L(k_1)^{1/3} P_L(k_2)^{2/3} P_L(k_3)}
{\alpha(k_2)^{1/3} \alpha(k_3)}  + 5 \; \rm{cyc.} \right] \nonumber\\
&& \;\;\; + c_3 \frac{ \left[ P_L(k_1) P_L(k_2) P_L(k_3) \right]^{2/3} }
{ \left[ \alpha(k_1) \alpha(k_2) \alpha(k_3) \right]^{1/3} } \Biggr \} .
\label{Bispectrum-delta-c}
\eeqa

Although the equilateral and orthogonal models are primarily defined via the shape of the bispectrum, 
they can also be characterized by the quadratic model (\ref{deltaL0-chi0-quadratic}).
From Eq.(\ref{bispectrum-deltaL0-fNL}), we can check that we recover the bispectrum 
(\ref{Bispectrum-delta-c}) with the choice
\beqa
&& \hspace{-0.3cm} f_{\rm NL}(\vk_1,\vk_2) = f_{\rm NL} \Biggl \{ 3 c_1 \frac{\alpha(k_3)}
{\alpha(k_1)\alpha(k_2)} \nonumber\\
&& \hspace{-0.3cm} + 3 c_2 \left[ \frac{\alpha(k_3)^{1/3} P_L(k_3)^{1/3}}
{\alpha(k_1) \alpha(k_2)^{1/3} P_L(k_2)^{1/3}} + \frac{\alpha(k_3)^{1/3} P_L(k_3)^{1/3}}
{\alpha(k_2) \alpha(k_1)^{1/3} P_L(k_1)^{1/3}} \right] \nonumber\\
&& \hspace{-0.3cm} + c_3 \frac{P_L(k_3)^{2/3}}
{\left[ \alpha(k_1) \alpha(k_2) \alpha(k_3) P_L(k_1) P_L(k_2) \right]^{1/3} } \Biggr \} ,
\label{eq:fNL_func_eq_ortho}
\eeqa
where $\vk_3=-\vk_1-\vk_2$.

In terms of the kernel $S_3$, we obtain from Eqs.(\ref{Bk-S3}) and (\ref{Bispectrum-delta-c}) the expression
\beqa
&& S_3(\vk_1,\vk_2,\vk_3) = f_{\rm NL} \Biggl \{ c_1 \left[ \frac{\alpha(k_3)}{\alpha(k_1) \alpha(k_2) P_L(k_3)} 
+ 2 \; \rm{cyc.} \right]
\nonumber\\
&& \;\;\; + c_2 \left[ \frac{\alpha(k_1)^{1/3}}{\alpha(k_2)^{1/3} \alpha(k_3) P_L(k_1)^{2/3} P_L(k_2)^{1/3}} 
+ 5 \; \rm{cyc.} \right]
\nonumber\\
&& \;\;\; + c_3 \frac{1}{\left[ \alpha(k_1) \alpha(k_2) \alpha(k_3) P_L(k_1) P_L(k_2) P_L(k_3) \right]^{1/3}}
\Biggr \} . 
\label{Bispectrum-S3-c}
\eeqa

\subsubsection{Squeezed limit}
\label{sec:squeezed}

In the squeezed limit, where one wave number goes to zero, the kernel $S_3$ of the local,
equilateral and orthogonal models defined by Eq.(\ref{Bispectrum-S3-c}) behaves as
\beqa
&& S_3(\vk',\vk,-\vk)_{k'\rightarrow 0} \sim f_{\rm NL} \left[ \frac{2 (c_1+c_2)}{\alpha(k') P_L(k)} \right.
\nonumber \\
&& \hspace{0.3cm} \left. +  \frac{2c_2+c_3}{\alpha(k')^{1/3} \alpha(k)^{2/3} P_L(k')^{1/3} P_L(k)^{2/3}} \right] .
\label{squeezed-S3-c}
\eeqa
For a large-scale power spectrum index $n_s$, $P_L(k) \propto k^{n_s}$, this gives
\beqa
&& k' \rightarrow 0 : \;\; P_L(k') S_3(\vk',\vk,-\vk)  \sim\,f_{\rm NL}
\nonumber \\
&& \times\,  \bigl [ 2(c_1+c_2) k'^{n_s-2} 
 + (2c_2+c_3) k'^{2(n_s-1)/3} \bigl ] ,
\label{PL-S3-kp0}
\eeqa
where we did not write the $k$-dependent dimensional factors.
Since $n_s \simeq 0.96$, we can see that both terms diverge at low $k'$, while higher-order
corrections vanish at low $k'$.
The divergence, as $(k')^{-1.04}$, is strongest for the local model where $c_1+c_2=1/3$. 
For the orthogonal model, where $c_1+c_2=0$, we only have the mild divergence
$(k')^{-0.03}$. 
For the equilateral model, where both $c_1+c_2=0$ and $2c_2+c_3=0$, the coefficients
of both divergent terms vanish. Thus, the equilateral model becomes Gaussian in the squeezed 
limit. 

As noted in Refs.~\cite{Scoccimarro:2011pz,Senatore:2009gt}, for the orthogonal model the template (\ref{Bispectrum-Phi-c}) is not very accurate in the squeezed limit and it gives a spurious divergence. Although Eq.~(\ref{Bispectrum-Phi-c}) may be used as a phenomenological model, if one wishes to study the orthogonal model physically defined from the inflationary models, one must use a specific model prediction or a more intricate template that regularizes the squeezed limit. Indeed, in the single-field inflation, the orthogonal shape is realized by a particular combination of the two types of cubic interaction terms \cite{Senatore:2009gt,Renaux-Petel2011,Renaux-Petel_etal2011}. But, the bispectrum generated by each interaction gives the same squeezed limit as in the case of the equilateral model. Hence, a physically-defined orhogonal shape leads to the vanishing first and second terms in Eq.~(\ref{PL-S3-kp0}), and one recovers Gaussianity in the squeezed limit. 

On the other hand, for models such as the local model (\ref{fNL-local}), where the expression (\ref{PL-S3-kp0}) diverges
or remains nonzero, we violate the condition (\ref{fNL-zero-mean}) associated
with the requirement $\langle \delta_L \rangle=0$. Indeed, to satisfy this constraint the expression
(\ref{PL-S3-kp0}) should vanish for $k' \rightarrow 0$.
In practice, for instance in numerical simulations, such infrared divergences are regularized 
by the finite size of the system. This means that there is no power on very large scales, beyond the simulation box,
and $P_L(k') S_3(\vk',\vk,-\vk)$ goes to zero at low $k'$. 
However, these models show large non-Gaussianities in the squeezed limit, on large cosmological scales 
below the cutoff, and they lead to different behaviors than the models where 
$P_L(k') S_3(\vk',\vk,-\vk)$ smoothly goes to zero without introducing an infrared cutoff.

\section{Correlation and response functions with primordial non-Gaussianities}
\label{sec:correlation_response}

Although the linear density field $\delta_{L0}$ is not Gaussian, we can apply 
with only modest modifications the approach that was described in 
\cite{Valageas2014a,Valageas2014b} to obtain consistency relations 
for cosmological structures
This relies on relations between correlation and response functions.
Thus, let us consider a set of quantities $\{\rho_i\}$ that are
functionals of $\delta_{L0}$, typically the nonlinear density contrasts
$\{\delta(\vk_i,\tau_i)\}$ at various wave numbers and conformal times,
and the mean response function $R(\vk')$ of their product with respect to the field 
$\delta_{L0}$ (which defines the initial conditions of the system),
\beq
R^{1,m}(\vk') = \langle \frac{\cD [ \rho_1 \dots  \rho_m ]}{\cD \delta_{L0}(\vk')} \rangle 
= \int \cD\delta_{L0} \; \cP(\delta_{L0}) 
\frac{\cD [ \rho_1 \dots  \rho_m ]}{\cD \delta_{L0}(\vk')} .
\label{R1n-def}
\eeq
Integrating by parts, we obtain
\beq
R^{1,m}(\vk') = - \int \cD\delta_{L0} \; \frac{\cD \cP}{\cD \delta_{L0}(\vk')} \; 
\rho_1 \dots  \rho_m .
\eeq
Using the expression (\ref{P-S3}), this yields at linear order over the kernels $S_n$
\beqa
R^{1,m}(\vk') & \!\! = \!\! & \langle \rho_1\dots \rho_m \, \delta_{L0}(-\vk') \rangle 
/ P_{\chi_0}(k') - \sum_{n=2}^{\infty} n \int \prod_{i=1}^{n-1} d\vk_i'  \nonumber \\
&& \times \delta_D(\vk'+\vk'_1+...+\vk'_{n-1}) S_n(\vk',\vk'_1,...,\vk'_{n-1}) \nonumber \\
&& \times \langle \rho_1\dots \rho_m \prod_{i=1}^{n-1} \delta_{L0}(\vk'_i) \rangle .
\label{R1n-1}
\eeqa
Note that the statistical average is with respect to the non-Gaussian distribution 
(\ref{P-S3}), but for the second term with the factors $S_n$ we can take the average
with the Gaussian weight only, as we work at linear order over $S_n$ or $f_{\rm NL}^{(n)}$.
Defining the mixed correlations $C^{\ell,m}(\vk'_1,\dots,\vk'_{\ell})$ as
\beq
C^{\ell,m}(\vk'_1,\dots,\vk'_{\ell}) = \langle \delta_{L0}(\vk'_1) \dots \delta_{L0}(\vk'_{\ell})
\rho_1\dots \rho_m \rangle ,
\label{Clm-def}
\eeq
Eq.(\ref{R1n-1}) reads as
\beqa
C^{1,m}(\vk') & = & P_{\chi_0}(k') R^{1,m}(-\vk') + P_{\chi_0}(k') \sum_{n=2}^{\infty}
n \int \prod_{i=1}^{n-1} d\vk'_i \nonumber \\
&& \times \; \delta_D\biggl( \vk' - \sum_{i=1}^{n-1} \vk'_i \biggl) 
S_n(-\vk',\vk'_1,...,\vk'_{n-1}) \nonumber\\
&& \times \; C^{n-1,m}(\vk'_1,...,\vk'_{n-1}) . \;\;\;
\label{C1m-R1m-Cn-1m}
\eeqa
We recover the relation between correlation and response functions associated with 
Gaussian initial conditions if we set $S_n=0$. Thus, the primordial non-Gaussianity 
introduces new terms that involve the higher-order mixed correlations
$C^{n-1,m}$.
Therefore, in the large-scale limit $\vk'\rightarrow 0$ we only recover the standard result
if these new terms go to zero, that is,
\beq
k' \rightarrow 0 : \;\;\; C^{1,m}(\vk') - P_{\chi_0}(k') R^{1,m}(-\vk') \rightarrow 0 , 
\label{C1m-kp0} 
\eeq
if we have
\beqa
&& \hspace{-0.5cm} \mbox{``squeezed Gaussianity''} : \;\;\; \mbox{for any} \;\;
\vk_1 + ... + \vk_n = 0 , \nonumber \\
&& P_{\chi_0}(0) \; S_{n+1}(0,\vk_1,...,\vk_n) = 0 . 
\label{C1m-kp0-condition1}
\eeqa
We can see that this condition is more stringent than the condition (\ref{Sn-norm-1})
associated with the constraint $\langle \delta_L \rangle=0$.
On the other hand, because it provides a natural way to satisfy Eq.(\ref{Sn-norm-1}), 
by making each term in that sum vanish, it defines a natural subclass of 
non-Gaussian models.

In particular, we can check that the generalized local models (\ref{Sn-def}) typically belong
to this class, as they obey Eq.(\ref{C1m-kp0-condition1}) and, in this manner,
satisfy Eq.(\ref{Sn-norm-1}).
Indeed, as we have already seen below Eq.(\ref{Sn-norm-1}), $P_{\chi_0}(0)=0$
and the term in $S_n$ with a factor $1/P_{\chi_0}(0)$ that appears in Eq.(\ref{Sn-def}) 
also vanishes because of the condition (\ref{fNL=0}).
However, as we have seen in section~\ref{sec:explicit} this is not the case in the explicit
local and orthogonal models, if they are defined by the template
(\ref{Bispectrum-Phi-c}). This is because of the inverse powers of $\alpha(k)$ that appear
in Eq.(\ref{Bispectrum-S3-c}) and lead to infrared divergences.
These divergences may be spurious and due to an inaccurate modeling of the squeezed limit
\cite{Senatore:2009gt}. In this case the condition (\ref{C1m-kp0-condition1}) is satisfied, once
we use a more accurate template, and we recover the Gaussian form (\ref{C1m-kp0-condition1}).
On the other hand, if the infrared divergence is meaningful, and only regularized by an additional
cutoff at very large scales, on cosmological scales below this cutoff the relations 
(\ref{C1m-R1m-Cn-1m}) show significant deviations from the Gaussian form.

From the expression (\ref{Bk-Sn}), we also find out that the property 
(\ref{C1m-kp0-condition1}) implies that the primordial density bispectrum 
vanishes in the squeezed limit,
\beq
\langle \delta_{L0}(0) \delta_{L0}(\vk) \delta_{L0}(-\vk) \rangle' = 0 .
\label{B-0-squeezed}
\eeq
This means that non-Gaussian models that have a nonzero squeezed bispectrum
violate the relationship (\ref{C1m-kp0}) between response and correlation functions.
On the other hand, a large class of non-Gaussian models, such as the generalized
local models (\ref{deltaL0-chi0}), satisfy Eq.(\ref{C1m-kp0-condition1}), which implies
that they recover the relationship (\ref{C1m-kp0}), which takes the same form as in the
Gaussian case, and they have a vanishing squeezed bispectrum 
(\ref{B-0-squeezed}).
From Eq.(\ref{Pk-Sn}), we find that these models also satisfy
\beq
k \rightarrow 0 : \;\;\; \frac{P_{L0}(k)}{P_{\chi_0}(k)} \rightarrow 1 .
\label{PL0-Pchi0-k0}
\eeq
Thus, in a broad sense, these models correspond to scenarios where the primordial
density field becomes Gaussian on very large scales, $k \rightarrow 0$,
or more precisely when at least one wave number goes to zero, in the squeezed limit.
This is why we may call the property (\ref{C1m-kp0-condition1}) as a squeezed
Gaussianity criterion.

Thus, we find that for models where the primordial bispectrum vanishes in the squeezed
limit (\ref{B-0-squeezed}) and non-Gaussianities are negligible in this limit, 
or more accurately where the condition (\ref{C1m-kp0-condition1})
is satisfied, the correlation $C^{1,m}$ in the large-scale limit $k'\rightarrow 0$ is set
by the gravitational dynamics (associated with the response $R^{1,m}$) as in the Gaussian
case.
Even though this relation takes the same form as in the Gaussian case,
it goes beyond the Gaussian model as it includes an implicit dependence on the 
primordial non-Gaussianity, because both quantities $C^{1,m}$ and $R^{1,m}$ 
depend on the properties of the initial conditions of the system, hence on the
kernels $S_n$.

\section{Consistency relations for the density contrast}
\label{sec:density_contrast}

\subsection{General case at all orders}
\label{sec:density-n-orders}

In practice, we consider the quantities $\{\rho_i\}$ to be nonlinear fields, such as the
nonlinear matter density contrasts $\delta(\vk_i,\tau_i)$.
As in the Gaussian case, they are fully determined by the linear field $\delta_{L0}$ 
that sets both the linear growing mode and the initial conditions 
(assuming as usual that decaying modes have had time 
to become negligible before gravitational clustering enters the nonlinear regime).
Then, we consider the mixed matter density correlation and response functions
\beq
C_{\delta}^{1,m}(\vk') = \langle \delta_{L0}(\vk') \delta(\vk_1,\tau_1) \dots 
\delta(\vk_m,\tau_m) \rangle ,
\label{C-delta-def}
\eeq
\beq
R_{\delta}^{1,m}(\vk') = \langle \frac{{\cal D} [ \prod_{j=1}^m \delta(\vk_j,\tau_j) ]}
{{\cal D}\delta_{L0}(\vk')} \rangle ,
\label{R-delta-def}
\eeq
and the relationship (\ref{C1m-R1m-Cn-1m}) reads as
\beqa
&& \hspace{-0.5cm} \langle \delta_{L0}(\vk') \prod_{j=1}^m \delta(\vk_j,\tau_j) \rangle
= P_{\chi_0}(k') \langle \frac{{\cal D} [ \prod_{j=1}^m \delta(\vk_j,\tau_j) ]}
{{\cal D}\delta_{L0}(-\vk')} \rangle \nonumber \\
&& + P_{\chi_0}(k') \sum_{n=2}^{\infty}
n \int \prod_{i=1}^{n-1} d\vk'_i \; \delta_D\biggl( \vk' - \sum_{i=1}^{n-1} \vk'_i \biggl) 
\nonumber \\
&& \times S_n(-\vk',\vk'_1,...,\vk'_{n-1}) \; 
\langle \prod_{i=1}^{n-1} \delta_{L0}(\vk'_i) \prod_{j=1}^m \delta(\vk_j,\tau_j) 
\rangle . \hspace{1cm}
\label{C1m-R1m-density}
\eeqa

As described in \cite{Valageas2014a}, in the limit of long-wavelength modes the
small-scale structures are transported by large-scale perturbations in a uniform 
fashion. This means that in the limit $k'\rightarrow 0$ for the support of a long-wavelength
perturbation $\Delta \delta_{L0}(\vk')$, the trajectories of the particles are simply
modified as
\beq
\vx(\vq,\tau) \rightarrow \vx(\vq,\tau) + D_+(\tau)  \Delta\Psi_{L0}(\vq) ,
\label{trajectories}
\eeq
where $\vq$ is the Lagrangian coordinate of the particles and $\Delta\Psi_{L0}(\vq)$,
which is uniform at leading order for $k' \rightarrow 0$, 
is the linear displacement field associated with the linear perturbation $\Delta \delta_{L0}$,
\beq
\Delta\Psi_{L0}(\vq) \equiv -\nabla_{\vq}^{-1} \cdot \Delta\delta_{L0} .
\label{DeltaPsiL0}
\eeq
The uniform shift (\ref{trajectories}) implies that the density field is modified as
\beq
\delta(\vx,\tau) \rightarrow \delta(\vx - D_+ \Delta\Psi_{L0} ,\tau) ,
\label{delta-x-transform}
\eeq
which reads in Fourier space (at linear order over $\Delta\Psi_{L0})$ as
\beq
\delta(\vk,\tau) \rightarrow \delta(\vk,\tau) - \ii D_+ (\vk\cdot\Delta\Psi_{L0}) \delta(\vk,\tau) .
\label{delta-k-transform}
\eeq
Then, one obtains
\beq
k' \rightarrow 0 : \;\;\; \frac{\cD \delta(\vk)}{\cD \delta_{L0}(\vk')} 
= D_+ \frac{\vk\cdot\vk'}{k'^2} \delta(\vk) ,
\label{Ddelta-DdeltaL0}
\eeq
\beq
R^{1,m}_{\delta}(\vk') = \langle \prod_{j=1}^m \delta(\vk_j,\tau_j) \rangle
\sum_{j=1}^m D_+(\tau_j) \frac{\vk_j\cdot\vk'}{k'^2} .
\label{R-kp-0}
\eeq
The results (\ref{Ddelta-DdeltaL0})-(\ref{R-kp-0}) follow from the weak equivalence
principle, that is, from the symmetries of the gravitational dynamics, and are independent
of the properties of the density field. Therefore, they remain valid for non-Gaussian initial
conditions, and Eq.(\ref{C1m-R1m-density}) becomes, in the squeezed limit, 
$k'\rightarrow 0$,
\beqa
&& \hspace{-0.5cm} \langle \delta_{L0}(\vk') \prod_{j=1}^m \delta(\vk_j,\tau_j) 
\rangle'_{k'\rightarrow 0} = - P_{\chi_0}(k') \; \langle \prod_{j=1}^m \delta(\vk_j,\tau_j)\rangle'
\nonumber \\
&& \times \sum_{j=1}^m D_+(\tau_j) \frac{\vk_j\cdot\vk'}{k'^2}
+ P_{\chi_0}(k') \sum_{n=2}^{\infty}
n \int \prod_{i=1}^{n-1} d\vk'_i \nonumber \\
&& \times \delta_D\biggl( \vk' - \sum_{i=1}^{n-1} \vk'_i \biggl) \;
 S_n(-\vk',\vk'_1,...,\vk'_{n-1}) \nonumber \\
&& \times \langle \prod_{i=1}^{n-1} \delta_{L0}(\vk'_i) 
\prod_{j=1}^m \delta(\vk_j,\tau_j) \rangle' . 
\label{C1m-R1m-density-kp0}
\eeqa

If the initial conditions obey the squeezed Gaussianity condition 
(\ref{C1m-kp0-condition1}) the relationship (\ref{C1m-R1m-density-kp0}) simplifies
and takes the same form as in the Gaussian case,
\beqa
&& \hspace{-0.3cm} \langle \delta_{L0}(\vk') \prod_{j=1}^m \delta(\vk_j,\tau_j) 
\rangle'_{k'\rightarrow 0} = - P_{L_0}(k') \; 
\langle \prod_{j=1}^m \delta(\vk_j,\tau_j)\rangle' \;\;\; \nonumber \\
&& \hspace{1cm} \times \sum_{j=1}^m D_+(\tau_j) \frac{\vk_j\cdot\vk'}{k'^2} .
\label{C1m-R1m-density-kp0-Gaussian0}
\eeqa
Using the fact that on large scales 
$\delta(\vk',\tau') \rightarrow D_+(\tau') \delta_{L0}(\vk')$, 
Eq.(\ref{C1m-R1m-density-kp0-Gaussian0}) also yields
\beqa
&& \hspace{-0.3cm} \langle \delta(\vk',\tau') \prod_{j=1}^m \delta(\vk_j,\tau_j) 
\rangle'_{k'\rightarrow 0} = - P_L(k',\tau') \; 
\langle \prod_{j=1}^m \delta(\vk_j,\tau_j)\rangle' \;\;\; \nonumber \\
&& \hspace{1cm} \times \sum_{j=1}^m \frac{D_+(\tau_j)}{D_+(\tau')} 
\frac{\vk_j\cdot\vk'}{k'^2} .
\label{C1m-R1m-density-kp0-Gaussian}
\eeqa

Thus, we can see that if the initial conditions show significant non-Gaussianities on
large scales and the coefficients (\ref{C1m-kp0-condition1}) do not vanish in the
squeezed limit, the consistency relation becomes much more complex than 
Eq.(\ref{C1m-R1m-density-kp0-Gaussian}).
The $(m+1)$-squeezed density correlation can no longer be expressed in terms of the
$m$-point small-scale density correlation, as there are additional contributions from
all-order mixed correlations (if all coefficients $S_n$ are nonzero). 
In particular, while the right-hand side in Eq.(\ref{C1m-R1m-density-kp0-Gaussian}) vanishes
at equal times as in the Gaussian case, which means that one must consider subleading
contributions, the new terms in Eq.(\ref{C1m-R1m-density-kp0}) do not vanish.

\subsection{Biased tracers}
\label{sec:Biased}

The consistency relations of the form (\ref{C1m-R1m-density-kp0-Gaussian}) in the
Gaussian case also apply to biased tracers  
\cite{Kehagias2013,Peloso2013,Peloso2014, Creminelli2013,Kehagias2014c,Creminelli2014a,Valageas2014a}.
Indeed, as recalled in section~\ref{sec:density-n-orders}, these consistency relations
follow from the weak equivalence principle, which states that all matter particles and
macroscopic objects fall at the same rate in a gravitational potential.
This means that under the almost uniform force $\nabla^{-1} \cdot \Delta\delta_{L0}$,
associated with the large-scale perturbation $\Delta\delta_{L0}$,
all particles and macroscopic objects experience the uniform shift (\ref{trajectories}).
Small-scale astrophysical processes, such as galaxy and star formation, 
are not modified by this uniform displacement so that all matter distributions,
including galaxy, cluster, and other biased tracers distributions, are simply shifted as in 
(\ref{delta-x-transform}). This is somewhat similar to the Galilean invariance of
usual hydrodynamical systems, and galaxies form and evolve following the
global flow of the system.
Here, because of the time-dependent cosmological
and gravitational setting, the uniform shift (\ref{trajectories}) involves the initial condition
and the linear growing mode $D_+(\tau)$.

Therefore, the consistency relation (\ref{C1m-R1m-density-kp0})
writes for biased tracers as
\beqa
&& \hspace{-0.5cm} \langle \delta(\vk',\tau') \prod_{j=1}^m \delta_g(\vk_j,\tau_j) 
\rangle'_{k'\rightarrow 0} = - P_{\chi}(k',\tau') \; \langle \prod_{j=1}^m \delta_g(\vk_j,\tau_j)\rangle'
\nonumber \\
&& \times \sum_{j=1}^m \frac{D_+(\tau_j)}{D_+(\tau')} \frac{\vk_j\cdot\vk'}{k'^2}
+ \frac{P_{\chi}(k',\tau')}{D_+(\tau')} \sum_{n=2}^{\infty}
n \int \prod_{i=1}^{n-1} d\vk'_i \nonumber \\
&& \times \delta_D\biggl( \vk' - \sum_{i=1}^{n-1} \vk'_i \biggl) \;
 S_n(-\vk',\vk'_1,...,\vk'_{n-1}) \nonumber \\
&& \times \langle \prod_{i=1}^{n-1} \delta_{L0}(\vk'_i) 
\prod_{j=1}^m \delta_g(\vk_j,\tau_j) \rangle' , 
\label{C1m-R1m-density-kp0-galaxies}
\eeqa
where $\delta_g$ is the density contrast of the biased tracers, such as galaxies.
Here we again used the large-scale asymptote, $\delta(\vk',\tau') \rightarrow D_+(\tau') \delta_{L0}(\vk')$, 
to replace $\delta_{L0}(\vk')$ by $\delta(\vk',\tau')$, where $\tau'$ is any arbitrary time.
Let us emphasize that Eq.(\ref{C1m-R1m-density-kp0-galaxies}) is exact and does
not make any assumption about the bias of the tracers.
In particular, it might be used to check the self-consistency or constrain phenomenological
models of biasing. 

Again, if the initial conditions obey the squeezed Gaussianity condition 
(\ref{C1m-kp0-condition1}), the relationship (\ref{C1m-R1m-density-kp0-galaxies}) simplifies
and takes the same form as in the Gaussian case,
\beqa
&& \hspace{-0.3cm} \langle \delta(\vk',\tau') \prod_{j=1}^m \delta_g(\vk_j,\tau_j) 
\rangle'_{k'\rightarrow 0} = - P(k',\tau') \langle \prod_{j=1}^m \delta_g(\vk_j,\tau_j)\rangle'  
\nonumber \\
&& \hspace{1cm} \times \sum_{j=1}^m \frac{D_+(\tau_j)}{D_+(\tau')} \frac{\vk_j\cdot\vk'}{k'^2} .
\label{C1m-R1m-density-kp0-Gaussian-galaxies0}
\eeqa

In practice, it could be convenient to write all density factors in the left-hand side of
Eq.(\ref{C1m-R1m-density-kp0-Gaussian-galaxies0}) in terms of the galaxy (or tracers)
density field, instead of the mixed matter-galaxy polyspectra.
On large scales we can expect a linear bias between the galaxy and matter
density fields,
\beq
k' \to 0 : \;\;\; \delta_g(\vk',\tau') = b_1(k',\tau') \, \delta(\vk',\tau') 
+ \epsilon (\vk',\tau') , 
\label{b1-def}
\eeq
where we included a stochastic component $\epsilon$. The latter represents shot noise
and the impact of small-scale nonlinear physics on galaxy formation. 
By definition of the split (\ref{b1-def}), it is uncorrelated with the
large-scale matter density field \cite{Hamaus2010}, 
$\langle \delta(\vk') \epsilon(\vk')\rangle'=0$.
We do not need to assume that the linear bias $b_1$ is scale independent.
In fact, it is well known that, for non-Gaussian initial conditions, the halo bias
can show a significant scale dependence. In particular, for the quadratic local model
(\ref{fNL-local}), the deviation of the bias from its Gaussian value 
behaves as $\Delta b_1 \propto f_{\rm NL} (b_1-1)/\alpha(k)$, which
diverges as $k^{-2}$ at low $k$ \cite{Dalal2008}.
This is related to the fact that such a model has strong non-Gaussianities on large scales
and does not verify the squeezed Gaussianity condition (\ref{C1m-kp0-condition1}).
We can expect the linear relationship (\ref{b1-def}) to hold for most cases,
for many astrophysical tracers and for both Gaussian and non-Gaussian initial
conditions. Both the matter and tracer density fluctuations $\delta^{(R)}(\vx)$ and
$\delta_g^{(R)}(\vx)$, smoothed over a large radius $R$, are small for sufficiently large
$R$, and Eq.(\ref{b1-def}) may be interpreted as the first-order term in a Taylor
expansion (we can also expect the different nonlocal terms associated with tidal effects 
to be small in this large-scale limit).
Then, Eq.(\ref{C1m-R1m-density-kp0-Gaussian-galaxies0}) simplifies (in the case of
squeezed Gaussianity) as
\beqa
&& \hspace{-0.3cm} \langle \delta_g(\vk',\tau') \prod_{j=1}^m 
\delta_g(\vk_j,\tau_j) \rangle'_{k'\rightarrow 0} = - b_1(k',\tau') P(k',\tau') \nonumber \\
&& \hspace{1cm} \times \langle \prod_{j=1}^m \delta_g(\vk_j,\tau_j)\rangle'  
\sum_{j=1}^m \frac{D_+(\tau_j)}{D_+(\tau')} \frac{\vk_j\cdot\vk'}{k'^2} .
\label{C1m-R1m-density-kp0-Gaussian-galaxies}
\eeqa
Here we assumed that the term $\langle \epsilon(\vk') \prod_{j=1}^m \delta_g(\vk_j) \rangle'$,
which appears when we replace $\delta(\vk')$ by Eq.(\ref{b1-def}), can be neglected.
This should be valid  in the regime where the consistency relations apply.
Indeed, in configuration space the consistency relations describe the effect of a large-scale
fluctuation $\delta(\vx')$ onto a distant small-scale region $\prod_j \delta_g(\vx_j)$,
which at leading order corresponds to the uniform displacement of the small-scale
structure towards (or away from) the large-scale region due to its gravitational attraction.
This means that, by assumption, the small-scale region is far away from the large-scale 
structure (or at least far from most of the large-scale region, simply because of the hierarchy
of scales) and only sensitive to its total mass at leading order. Then, the stochastic component 
$\epsilon(\vk')$ within the large-scale region, which is decorrelated from the large-scale mass 
by definition ($\langle \epsilon \delta \rangle=0$) and corresponds to local processes inside
this region, should be independent or very weakly correlated with the galaxy density field
in the distant small-scale region. This implies that it must be negligible as compared
with the right-hand side in Eq.(\ref{C1m-R1m-density-kp0-Gaussian-galaxies}),
which behaves as $b_1(k') (k')^{n_s-1}$, with at worse a very slow decrease at low $k'$ for 
$n_s \simeq 0.96$ if $b_1$ does not vanish, whereas we can expect 
$\langle \epsilon(\vk') \prod_{j=1}^m \delta_g(\vk_j) \rangle'$ to decrease much faster at low $k'$.
However, in Eq.(\ref{C1m-R1m-density-kp0-Gaussian-galaxies}) we cannot replace the
prefactor $b_1 P(k')$ by the galaxy power spectrum, $P_g(k')/b_1$, because 
from Eq.(\ref{b1-def}) we have $P_g = b_1^2 P + P_{\epsilon}$ and the shot noise or stochastic
power spectrum cannot be neglected with respect to the matter power spectrum on large scales.

\subsection{Bispectrum}
\label{sec:Bispectrum}

The simplest density consistency relation is obtained for the squeezed bispectrum,
with $m=2$. 
We consider the case of biased tracers for the sake of generality,
in the general and exact form (\ref{C1m-R1m-density-kp0-galaxies}) that does not
explicitly involve the bias and makes no assumption on the biasing of the tracers.
The case of the matter density contrast $\delta$ is recovered by
replacing $\delta_g(\vk)$ and $P_g(k)$ by $\delta(\vk)$ and $P(k)$

\subsubsection{Scenarios with squeezed Gaussianity}
\label{sec:LS-G-bispectrum}

For models that satisfy the squeezed Gaussianity criterion 
(\ref{C1m-kp0-condition1}), Eq.(\ref{C1m-R1m-density-kp0-Gaussian-galaxies0}) gives the
usual result
\beqa
&& \hspace{-0.3cm} \langle \delta(\vk',\tau') \delta_g(\vk_1,\tau_1) \delta_g(\vk_2,\tau_2) 
\rangle'_{k'\rightarrow 0} = - P(k',\tau') \nonumber \\
&& \hspace{0cm} \times P_g(k_1;\tau_1,\tau_2)
\left[ \frac{D_+(\tau_1)}{D_+(\tau')} \frac{\vk_1\cdot\vk'}{k'^2} 
+ \frac{D_+(\tau_2)}{D_+(\tau')} \frac{\vk_2\cdot\vk'}{k'^2} \right] . \hspace{0.7cm}
\label{bispectrum-kp-Gaussian}
\eeqa
This takes the same form as the consistency relation obtained for Gaussian initial 
conditions, but it includes the effect of primordial non-Gaussianities (up to first order
over the kernels $S_n$) as the bispectrum in the left-hand side and the nonlinear
power spectrum $P_g(k_1;\tau_1,\tau_2)$ in the right-hand side are sensitive to these
primordial non-Gaussianities.
As in the Gaussian case, this relation vanishes at equal times, which means that
the equal-time bispectrum is governed by the subleading contributions
\cite{Valageas2014b,Nishimichi2014}.

If the bias is linear on large scales, as in Eq.(\ref{b1-def}),
we can replace $\delta(\vk')$ by $\delta_g(\vk')/b_1$ as was done in 
Eq.(\ref{C1m-R1m-density-kp0-Gaussian-galaxies}), the stochastic contribution being
subdominant.

\subsubsection{$S_3$-type primordial non-Gaussianity}
\label{sec:S3-bispectrum}

In the case of the $S_3$-type models introduced in section~\ref{sec:quadratic},
where only the kernel $S_3$ is nonzero, the relation (\ref{C1m-R1m-density-kp0-galaxies}) 
gives
\beqa
&& \hspace{-0.3cm} \langle \delta(\vk',\tau') \delta_g(\vk_1,\tau_1) \delta_g(\vk_2,\tau_2) 
\rangle'_{k'\rightarrow 0} = - P(k',\tau') \nonumber \\
&& \hspace{0cm} \times P_g(k_1;\tau_1,\tau_2)
\left[ \frac{D_+(\tau_1)}{D_+(\tau')} \frac{\vk_1\cdot\vk'}{k'^2} 
+ \frac{D_+(\tau_2)}{D_+(\tau')} \frac{\vk_2\cdot\vk'}{k'^2} \right] \nonumber \\
&& + 3 \frac{P(k',\tau')}{D_+(\tau')} \int d\vk'_1 d\vk'_2 \;
\delta_D(\vk' - \vk'_1 - \vk'_2) \nonumber \\
&& \times S_3(-\vk',\vk'_1,\vk'_2) \; \langle \delta_{L0}(\vk'_1)  \delta_{L0}(\vk'_2) 
\delta_g(\vk_1,\tau_1) \delta_g(\vk_2,\tau_2) \rangle' . \nonumber \\
&&
\label{S3-bispectrum-density-kp0}
\eeqa
Here we used the result (\ref{PL0-S3}) that $P_{\chi_0}=P_{L0}$ for these models.
As compared with the Gaussian case, there is an additional contribution to the
consistency relation involving the mixed four-point correlation.

At equal times, the first term in the right-hand side of 
Eq.(\ref{S3-bispectrum-density-kp0}) again vanishes. This is because it arises from the
uniform displacement of the small-scale structures by the long-wavelength mode $k'$
\cite{Valageas2014a},
which cannot be detected from the properties of the equal-time density field.
However, this term is only the leading-order factor associated with the gravitational 
dynamics.
It scales as $1/k'$ [multiplied by the prefactor $P(k')$], and there are subleading 
corrections that remain finite at low $k'$ and do not vanish at equal times 
\cite{Valageas2014b}.
However, the contributions from these latter terms go to zero in the limit $k' \rightarrow 0$
because of the prefactor $P(k') \rightarrow 0$.
Then, if the primordial non-Gaussianities are sufficiently high on large scales,
the right-hand side will be dominated by the second term. This applies for instance
to the usual local model (\ref{local-Phi}), where $S_3(-\vk',\vk'_1,\vk'_2)$ diverges
for $k' \rightarrow 0$.
Then, we obtain for such models at equal times,
\beqa
&& \hspace{-0.3cm} \langle \delta(\vk') \delta_g(\vk_1) \delta_g(\vk_2) 
\rangle'_{k'\rightarrow 0} = 
3 \frac{P(k')}{D_+} \int d\vk'_1 d\vk'_2 \nonumber \\
&& \times \; \delta_D(\vk' - \vk'_1 - \vk'_2) \; S_3(-\vk',\vk'_1,\vk'_2) \nonumber \\
&& \times \langle \delta_{L0}(\vk'_1)  \delta_{L0}(\vk'_2) 
\delta_g(\vk_1) \delta_g(\vk_2) \rangle' .
\label{S3-bispectrum-density-equal}
\eeqa
If we split the four-point function in the right-hand side of 
Eq.(\ref{S3-bispectrum-density-equal}) over connected and disconnected parts,
we obtain
\beqa
&& \hspace{-0.5cm} \langle \delta(\vk') \delta_g(\vk_1) \delta_g(\vk_2) 
\rangle'_{k'\rightarrow 0} = 6 \frac{P(k')}{D_+} P_{L0,g}(k_1) P_{L0,g}(k_2) 
\nonumber \\
&& \times S_3(\vk',\vk_1,\vk_2) + 3 \frac{P(k')}{D_+} \int d\vk'_1 d\vk'_2 \; 
\delta_D(\vk' - \vk'_1 - \vk'_2) \nonumber \\
&& \times S_3(-\vk',\vk'_1,\vk'_2) \;
\langle \delta_{L0}(\vk'_1)  \delta_{L0}(\vk'_2) \delta_g(\vk_1) \delta_g(\vk_2) \rangle'_c ,
\label{S3-bispectrum-density-disconnected}
\eeqa
where the subscript ``c'' in the last term $\langle \dots \rangle'_c$ denotes the
connected part, that is, the trispectrum, and 
$P_{L0,g} = \langle \delta_{L0} \delta_g \rangle'$ is the mixed linear-nonlinear
power spectrum.
On large scales, where at zeroth order over $S_3$ we have $\delta_L=\chi$
and the trispectrum vanishes, we recover the expression (\ref{Bk-S3})
(for unbiased tracers or for the matter density contrast), that is,
the primordial bispectrum of the linear density field is set by the initial non-Gaussianity.
On smaller scales, where $|\vk_1| = |\vk_2|$ enters the nonlinear regime,
we can see that the first term in the right-hand side of 
Eq.(\ref{S3-bispectrum-density-disconnected}) keeps the same form, except that 
the linear power spectra $P_{L0}(k_1)$ and $P_{L0}(k_2)$ are replaced by the mixed
linear-nonlinear power spectra $P_{L0,g}(k_1)$ and $P_{L0,g}(k_2)$,
while the trispectrum contribution no longer vanishes.
In practice, this means that it is difficult to use the relationship 
(\ref{S3-bispectrum-density-disconnected}) for analytical or observational purposes,
because the mixed trispectrum 
$\langle \delta_{L1'} \delta_{L2'} \delta_1 \delta_2 \rangle'$ is not easier to
model or compute than the bispectrum $\langle \delta' \delta_1 \delta_2 \rangle'$
and it cannot be directly observed.
If the kernel $S_3(-\vk',\vk_1',\vk_2')$ in the second term peaks at low values of
$k_1'$ and $k_2'$ that are still in the linear regime, we can replace the mixed
trispectrum by the nonlinear trispectrum, as $\delta_L(\vk'_1) \simeq \delta(\vk_1)$
and $\delta_L(\vk'_2) \simeq \delta(\vk_2)$.
This trispectrum can in principle be measured, so that 
Eq.(\ref{S3-bispectrum-density-disconnected})
could be compared with observations (or simulations).
However, because trispectra are usually very noisy and difficult to measure,
it is unlikely that the relationship (\ref{S3-bispectrum-density-disconnected})
will provide a competitive method to probe such non-Gaussian scenarios.

\subsubsection{Explicit examples of primordial non-Gaussianity}
\label{sec:bispectrum-explicit}

Let us examine the specific models of primordial non-Gaussianity introduced in 
Sec.~\ref{sec:explicit} to see whether Eqs.~(\ref{S3-bispectrum-density-equal}) or 
(\ref{S3-bispectrum-density-disconnected}) receive a nonvanishing correction or not. 
This is set by the behavior at low $k'$ of the expression (\ref{PL-S3-kp0}).
As noticed in section~\ref{sec:squeezed}, for the models defined by the template 
(\ref{Bispectrum-Phi-c}), the expression (\ref{PL-S3-kp0}) diverges at low $k'$, along with the bispectrum in the squeezed limit, except for the equilateral model where the coefficients
of the two divergent terms vanish.
The divergence is strongest for the local model (\ref{fNL-local}).
Then, the standard consistency relation is largely violated. 
In contrast, for the equilateral model, and the orthogonal model defined by
physical inflation scenarios instead of the template (\ref{Bispectrum-Phi-c}),
we recover the standard consistency relation.
In general, models with a large squeezed bispectrum lead to a nonvanishing equal-time 
correlation. Thus, apart from a difficulty in predicting the size of these corrections, 
Eq.~(\ref{S3-bispectrum-density-disconnected}) has a sizable effect for primordial 
non-Gaussianities with a large squeezed bispectrum, and this relation may be used as a 
consistency check of other non-Gaussian probes.

\subsection{Lack of constraints on equal-times bias models}
\label{sec:local-bias}

We may note that for local bias models, where we write expansions such as
$\delta_g = b_1 \delta + b_2 \delta^2/2 + ...$,
the galaxy power spectrum and the matter-galaxy-galaxy bispectrum read up to linear order over $b_2$
as
\beqa
P_g(k) & = & b_1^2 P(k) + b_1 b_2 \int d\vk_1' d\vk_2' \delta_D(\vk_1'+\vk_2'-\vk) \nonumber \\
&& \times \langle \delta(\vk_1') \delta(\vk_2') \delta(-\vk) \rangle' ,
\label{bispectrum-kp-Gaussian-bias-0}
\eeqa
and
\beqa
&& \hspace{-0.5cm} \langle \delta(\vk') \delta_g(\vk_1) \delta_g(\vk_2) \rangle' = 
b_1^2 \langle \delta(\vk') \delta(\vk_1) \delta(\vk_2) \rangle' \nonumber \\
&& \hspace{0cm} + b_1 b_2 P(k') \left[ P(k_1) + P(k_2) \right] + \frac{b_1 b_2}{2} \int d\vk_1' d\vk_2' \nonumber \\
&& \times \bigl [ \delta_D(\vk_1'+\vk_2'-\vk_1) \langle \delta(\vk') \delta(\vk_2) \delta(\vk_1') \delta(\vk_2') \rangle'_c
\nonumber \\
&& + \delta_D(\vk_1'+\vk_2'-\vk_2) \langle \delta(\vk') \delta(\vk_1) \delta(\vk_1') \delta(\vk_2') \rangle'_c \bigl ] .
\label{bispectrum-kp-Gaussian-bias}
\eeqa

We may compare with Eq.(\ref{bispectrum-kp-Gaussian}), obtained for Gaussian
or squeezed-Gaussianity initial conditions.
These two expressions are actually quite different.
The expression (\ref{bispectrum-kp-Gaussian-bias}) is usually taken for equal times
and it explicitly involves the linear and quadratic bias parameters $b_1$ and $b_2$.
In contrast, the relation (\ref{bispectrum-kp-Gaussian})
vanishes at equal times and it does not explicitly involve $b_1$ and $b_2$.
Nevertheless, these two relations are not contradictory at equal times. 
Indeed, the consistency relation (\ref{bispectrum-kp-Gaussian}) only gives the leading-order
contribution at unequal times, which is actually much greater than the second term in
Eq.(\ref{bispectrum-kp-Gaussian-bias}) by a factor $1/k'$, but does not give the
explicit expression of the equal-times bispectrum, although its vanishing at equal times 
implies that the equal-times bispectrum must grow more slowly than $P_L(k')/k'$ at
low $k'$, which is satisfied by Eq.(\ref{bispectrum-kp-Gaussian-bias}).
Therefore, Eq.(\ref{bispectrum-kp-Gaussian-bias}) at equal times does not contradict
Eq.(\ref{bispectrum-kp-Gaussian}) and it involves higher-order contributions that are
not taken into account in this consistency relation.

Ref.~\cite{Kehagias2014c} explicitly checked that a quadratic local bias model satisfies the consistency
relation for the galaxy bispectrum, up to one-loop order. In fact, we can see that any bias model,
where the galaxy density field is a functional of the same-time matter density field, automatically satisfies
the consistency relations (\ref{C1m-R1m-density-kp0-galaxies}), for both Gaussian and non-Gaussian
initial conditions and at all orders.
Indeed, let us write the galaxy density field as an expansion over the matter density field,
\beqa
\delta_g(\vk,\tau) & = & \sum_{\ell=0}^{\infty} \int d\vk_1 \dots d\vk_{\ell} \;
\delta_D(\vk_1+\dots+\vk_{\ell} - \vk) \nonumber \\
&& \times b_{\ell}(\vk_1,\dots,\vk_{\ell} ; \tau) \delta(\vk_1,\tau) \dots \delta(\vk_{\ell},\tau) .
\label{bias-full}
\eeqa
This expression goes beyond the local bias models, because the $\vk$ dependence of the kernels $b_\ell$
allows us to include tidal terms, which are naturally generated by the dynamics 
\cite{Chan2012,Baldauf2012,McDonald2009}.
We can also include stochasticity as the kernels $b_\ell$ can be stochastic, so that after the average
$\langle\dots\rangle$ over the initial conditions we perform a second average
over some stochastic variables $\epsilon_i$ that are uncorrelated with the density field.
Thus, Eq.(\ref{bias-full}) is a very general bias model, which only assumes that the galaxy density field
can be expanded over powers of the same-time matter density field.
Then, substituting Eq.(\ref{bias-full}) into the left-hand side in Eq.(\ref{C1m-R1m-density-kp0-galaxies})
and using the matter consistency relation (\ref{C1m-R1m-density-kp0}), we recover the right-hand
side in Eq.(\ref{C1m-R1m-density-kp0-galaxies}).
This is straightforward for the non-Gaussian $S_n$-dependent term in Eq.(\ref{C1m-R1m-density-kp0-galaxies}),
and it also applies to the first standard term because each factor of the form
$(\vk_1^{(j)} + \dots + \vk_{\ell_j}^{(j)})\cdot\vk'/k'^2$, associated with the expansion
(\ref{bias-full}) of a factor $\delta_g(\vk_j)$, simplifies as $\vk_j \cdot\vk'/k'^2$ thanks to the Dirac factors
in Eq.(\ref{bias-full}).
Therefore, we find that the consistency relations (\ref{C1m-R1m-density-kp0-galaxies})
do not provide very useful constraints or guidelines for the building of analytic biasing models, as they
do not give any information on the kernels $b_{\ell}$ and all bias models (\ref{bias-full}) satisfy these
consistency relations.
On the other hand, this ensures that the general bias models (\ref{bias-full}) do not face unphysical
inconsistencies at this level.

As noticed in Ref.~\cite{Kehagias2014c}, this result is expected because the bias model (\ref{bias-full})
verifies the symmetries of the system. More explicitly, the galaxy density field defined by
Eq.(\ref{bias-full}) is transported in a uniform fashion by a large-scale mode, exactly as in 
Eq.(\ref{delta-x-transform}) for the matter density field, as all factors $\delta(\vx_i,\tau)$
(working in configuration space) are shifted by the same uniform displacement $D_+(\tau)  \Delta\Psi_{L0}$ 
as in Eq.(\ref{trajectories}).
This implies that $\delta_g$ also verifies the response (\ref{Ddelta-DdeltaL0}) to a large-scale
perturbations, which directly leads to the consistency relations.

On the other hand, if we consider a bias model that involves unequal-times matter density fields 
$\delta(\vk_i,\tau_i)$ in Eq.(\ref{bias-full}), with kernels $b_{\ell}(\vk_1,\tau_1;...;\vk_{\ell},\tau_{\ell})$ 
and integrals over the past times $\tau_i$, the consistency relations are generically violated.
Indeed, the different factors $\delta(\vx_i,\tau_i)$ (working again in configuration space) are shifted by the
different uniform displacements $D_+(\tau_i)  \Delta\Psi_{L0}$.
In Fourier space, we can no longer use the simplification 
$(\vk_1^{(j)} + \dots + \vk_{\ell_j}^{(j)})\cdot\vk'/k'^2 = \vk_j \cdot\vk'/k'^2$, because each factor
$\vk_i^{(j)} \cdot\vk'/k'^2$ is multiplied by a different time-dependent factor $D_+(\tau_i^{(j)})$.
This means that the consistency relations provide strong constraints on bias models that write the
galaxy density field as a functional of different-times matter density fields.
However, in practice bias models have the equal-time form (\ref{bias-full}), to avoid unnecessarily
complex models that display too many free parameters and free functions, and to focus on the
simplest models.

\section{Consistency relations  for velocity and momentum fields}
\label{sec:consistency_rel-v-p}

As we recalled in the previous section, the leading-order effect of a long-wavelength 
perturbation is to move smaller structures by uniform shift, which leads to the
functional derivative (\ref{Ddelta-DdeltaL0}). 
Then, if we consider equal-time statistics of the density field, in the Gaussian
case we cannot see any effect (as we cannot detect a uniform shift by such probes)
and the sum in the right-hand side of Eq.(\ref{C1m-R1m-density-kp0-Gaussian0}) 
vanishes (using $\sum \vk_j=0$). 
On the other hand, for scenarios with high primordial non-Gaussianities on large scales,
the new term in the right-hand side of Eq.(\ref{C1m-R1m-density-kp0})
remains significant in the squeezed limit and we obtain a nonzero relationship,
as in Eq.(\ref{S3-bispectrum-density-disconnected}).

As pointed out in \cite{Rizzo2016a}, in the Gaussian case we can obtain
nontrivial consistency relations by cross correlating density and velocity, or momentum,
fields. Indeed, the uniform displacement of the small-scale structures also leads
to a modification of the amplitude of the local velocity and the latter can be detected
at equal times by measuring the velocity or momentum field.
In a fashion similar to the procedure described in section~\ref{sec:density-n-orders}
for density correlations, we can obtain these consistency relations from the general
relationship (\ref{C1m-R1m-Cn-1m}) by taking the quantities $\{\rho_i\}$ to be
a combination of density and velocity fields. 
For instance, considering the momentum field $\vp(\vx,\tau)$ defined by
\beq
\vp = (1+\delta) \vv ,
\label{p-def}
\eeq
we obtain at equal times the nonzero consistency relation
\beqa
&& \langle \delta(\vk') \prod_{j=1}^m \delta(\vk_j) \!
\prod_{j=m+1}^{m+\ell} \vp(\vk_j) \rangle_{k'\rightarrow 0}'  = - \ii \, P_L(k') 
\frac{d\ln D_+}{d\tau} \nonumber \\
&& \times \sum_{i=m+1}^{m+\ell} \langle \prod_{j=1}^m \! \delta(\vk_j) \!
\prod_{j=m+1}^{i-1} \! \vp(\vk_j) \left( \frac{\vk'}{k'^2} [ \delta_D(\vk_i) + \delta(\vk_i) ] \right)  
\! \nonumber \\
&& \times \prod_{j=i+1}^{m+\ell} \!\! \vp(\vk_j)  \rangle' ,
\label{consistency_relation_p-single-time}
\eeqa
where we did not write the common time $\tau$ of all fields.
To obtain scalar consistency relations, instead of the vector quantities 
(\ref{consistency_relation_p-single-time}), we can consider the
divergence of the momentum field, 
\beq
\lambda(\vx,\tau) \equiv \nabla_{\vx} \cdot \left[ (1+ \delta) \vv \right] , \;\;\;
\lambda(\vk,\tau) \equiv \ii \vk \cdot \vp(\vk,\tau) .
\label{lambda-def}
\eeq
Then, the consistency relations for the divergence $\lambda$ follow from
those obtained for $\vp$ and read at equal times as
\beqa
&& \hspace{-1cm} \langle \delta(\vk') \prod_{j=1}^m \delta(\vk_j) 
\prod_{j=m+1}^{m+\ell} \lambda(\vk_j) \rangle_{k'\rightarrow 0}'  = P_L(k') 
\frac{d \ln D_+}{d\tau} \nonumber \\
&& \hspace{-0.2cm} \times \sum_{i=m+1}^{m+\ell} \frac{\vk_i\cdot\vk'}{k'^2} 
\langle \delta(\vk_i) 
\prod_{j=1}^m \delta(\vk_j) \prod_{\substack{j=m+1 \\ j\neq i}}^{m+\ell} \lambda(\vk_j) 
\rangle' , \;\;\;
\label{consistency_relation_lambda-single-time}
\eeqa
which is again nonzero.

The simplest consistency relation that does not
vanish at equal times is the equal-time bispectrum with one momentum field. 
From Eqs.(\ref{consistency_relation_p-single-time}) and 
(\ref{consistency_relation_lambda-single-time}), we obtain
\beq
\langle \delta(\vk') \delta(\vk) \vp(-\vk) \rangle_{k'\rightarrow 0}'  
= - \ii \frac{\vk'}{k'^2} \frac{d\ln D_+}{d\tau} P_L(k') P(k) \;\;\;
\label{bispectrum_p}
\eeq
and
\beq 
\langle \delta(\vk') \delta(\vk) \lambda(-\vk) \rangle_{k'\rightarrow 0}'  
= -\frac{\vk\cdot\vk'}{k'^2} \frac{d\ln D_+}{d\tau} P_L(k') P(k) . \;\;\;
\label{bispectrum_lambda}
\eeq
Here $P(k)$ is the nonlinear density power spectrum and these
relations remain valid in the nonperturbative nonlinear regime.

These momentum consistency relations were derived in \cite{Rizzo2016a} for the Gaussian case.
As for the density contrast studied in section~\ref{sec:density_contrast},
for non-Gaussian models these consistency relations are modified by the additional terms 
that arise from the $S_n$-dependent factors in Eq.(\ref{C1m-R1m-Cn-1m}), 
in a fashion similar to Eq.(\ref{C1m-R1m-density-kp0}).
We do not explicitely write these terms here, because they lead to consistency relations
that are probably of little practical value since they involve high-order
mixed correlations $\langle \prod \delta_L \prod \delta \prod \vv \rangle'$
or $\langle \prod \delta_L \prod \delta \prod \vp \rangle'$
that are difficult to measure or predict.
On the other hand, for non-Gaussian scenarios that obey the squeezed Gaussianity conditions 
(\ref{C1m-kp0-condition1}) the consistency relations 
(\ref{consistency_relation_p-single-time})-(\ref{bispectrum_lambda}) remain valid, as 
Eq.(\ref{C1m-R1m-density-kp0-Gaussian}) for the density contrast.
Again, even if these relations now take the same form as in the Gaussian case, they go beyond
the Gaussian result because the correlation functions in both sides of these relations depend
on the properties of the initial conditions, hence on the kernels $S_n$.

\section{Conclusions}
\label{sec:Conclusions}

In this paper, we have described how the consistency relations of cosmological
large-scale structures are modified when the initial conditions are not Gaussian.
We consider very general scenarios, where the primordial density field can be written
as a nonlinear functional of a Gaussian field $\chi_0$, up to all orders over $\chi_0$,
or more generally, where the probability distribution of the primordial density field 
can be expanded around the Gaussian, up to all orders over $\delta_{L0}$.
We also give the relationship between these two formalisms.
As the primordial non-Gaussianities should be small, to remain consistent with
observations, we work at linear order over the non-Gaussianity kernels 
$f_{\rm NL}^{(n)}$ or $S_n$.
We give the constraints that must be verified by these kernels, which arise from the 
normalization conditions $\langle 1 \rangle =1$ and $\langle \delta_{L0} \rangle =0$.
A simple example is provided by the $S_3$-type primordial non-Gaussianity,
which is fully defined by the power spectrum and bispectrum.

We show how the approach used for the Gaussian case applies to these
non-Gaussian scenarios. We can still obtain a relationship between correlation
and response functions, but it is generally much more intricate as it involves
all-order mixed correlations such as $\langle \prod \delta_L \prod \delta \rangle'$.
For scenarios that converge to the Gaussian in the squeezed limit, we recover the
simple relationship obtained in the Gaussian case, even though the small-scale
modes may be strongly affected by the primordial non-Gaussianity.
We give the explicit conditions for this simplification to hold.

Then, as for the Gaussian case, we use this general relationship to derive the
consistency relations for density and momentum fields, as well as for biased tracers.
We discuss in more details the relation obtained for the density bispectrum,
especially for the case of the $S_3$-type primordial non-Gaussianity.
We describe the form it takes at equal times, when the Gaussian-like term vanishes
and we are dominated by the new contributions associated with the primordial
non-Gaussianity.
Unfortunately, this expression involves a complicated mixed trispectrum
$\langle \delta_L\delta_L \delta\delta \rangle'$ and it may not be very practical.
In the case of scenarios with squeezed Gaussianity, 
we briefly discuss the relations obtained for the bispectrum with one momentum
field, as they remain nonzero at equal times.

We find that, for both Gaussian and non-Gaussian initial conditions, the consistency relations
are automatically satisfied by very general bias models, where the galaxy density field can be
written as an expansion over powers of the same-time density field (including nonlocal terms such as
tidal fields and stochasticity). Thus, these consistency relations do not provide any information
on the generalized bias kernels $b_{\ell}(\vk_1,...,\vk_{\ell};\tau)$.
On the other hand, this means that such bias models do not face unphysical inconsistencies at this level. 
In contrast, the consistency relations would constrain bias models that write the galaxy density field
as a functional of different-times density fields, but such models are not used in practice.

To conclude, we find that the usual consistency relations, that were derived for
Gaussian initial conditions, remain valid for a large class of primordial non-Gaussianities.
Even though the small-scale modes probed by these relations may be highly nonlinear
or strongly affected by the primordial non-Gaussianity, it is sufficient for their validity
that primordial non-Gaussianities vanish in the squeezed limit, when one mode
is pushed to large scales.
Therefore, these consistency relations cannot be used as a precise test of such scenarios,
in the sense that they do not discriminate between the Gaussian case and these models.
On the other hand, this means that within this general class of primordial fluctuations
the consistency relations remain a test of deviations from General Relativity, which is the
other hypothesis used in their derivation (more precisely, the weak equivalence principle).

If the primordial non-Gaussianities remain large in the squeezed limit,
the consistency relations are modified and involve additional mixed linear-nonlinear 
correlations $\langle \prod \delta_L \prod \delta \rangle'$.
Then, cosmological consistency relations can be used as a test of such
squeezed primordial non-Gaussianities.
A lack of deviation from the Gaussian-case prediction (derived within General Relativity) 
would constrain the amplitude of squeezed primordial non-Gaussianities and of deviations
from General Relativity, whereas a measured deviation would either rule out both the Gaussian 
scenario and the non-Gaussian models with vanishing squeezed primordial non-Gaussianities,
or provide evidence for deviations from General Relativity.

These consistency relations should be considered as a null test of General Relativity
and Gaussian initial conditions, which we have shown in this paper must be extended
to the broader class of squeezed Gaussianity. They do not really provide a new probe
to measure primordial non-Gaussianities (if they are nonzero), as they rely on the measurement 
of the bispectrum (or higher-order correlations), which by itself is already a standard probe of 
primordial non-Gaussianities.

\begin{acknowledgments}

This work is supported in part by the French Agence Nationale de la Recherche
under Grant ANR-12-BS05-0002 (PV), and MEXT/JSPS KAKENHI Grant Number JP15H05899 
and JP16H03977 (AT). 

\end{acknowledgments}

\bibliography{ref1}   

\end{document}